\documentstyle[aps,psfig,multicol,twoside]{revtex}

\begin{document}

\voffset      = 0.2truecm
\textheight   =24.7truecm
\baselineskip = 0.4truecm
\headsep      = 1.0truecm

\title{Reply to ``Comment on `Pipe network model for 
scaling of dynamic interfaces in porous media' ''}

\author{Chi-Hang Lam$^{1}$ and Viktor K. Horv\'{a}th$^2$} 

\address{$^1$Department of Applied Physics, Hong Kong Polytechnic
University, Hung Hom, Hong Kong\\ $^2$Department of Physics,
University of Pittsburgh, Pittsburgh, Pennsylvania 15260, and\\ 
Dept. of Biological Physics, E\"{o}tv\"{o}s Univ., P\'azm\'any
P. 1A. 1117 Budapest, Hungary}

\date{\today}

\maketitle

\pacs{PACS numbers: 05.40.+j, 47.55.Mh, 05.70.Ln, 68.35.Fx}


\newsavebox{\boxl}
\savebox{\boxl}{$\ell$}
\begin{multicols}{2}
In a previous experiment \cite{Horvath}, imbibition fronts moving up
in a paper sheet, that drifts down into a water tank at speed $v$,
were found to have an average height $\bar{h} \sim v^{-1/\Omega}$, a
width $w \sim \bar{h}^{\Omega\kappa}$, and a temporal correlation
function $C(t) = v^{-\kappa} f (t\/v^{(\theta_{t}+\kappa)/\beta})$.
We simulated the phenomenon using a pipe network model and reproduced
the rich scaling behaviors with $all$ exponents $\Omega$, $\kappa$,
$\theta_t$ and $\beta$ in agreement with the experimental values
\cite{Lam}. 
Another study based on a phase-field model leads to predictions on
exponents that are unfortunately incompatible with the
experimental ones\cite{Dube}.

The front propagation exemplifies two related dynamics, namely mean
flow and roughening. Assuming a unique time scale controlling both
dynamics, an exponent identity $\beta =
\Omega(\theta_t+\kappa)/ (\Omega+1)$ was derived. Inserting
experimental values into the right-hand side, we get $\beta=0.52$
compared with the measured value 0.56. Using numerical estimates,
$\beta = 0.54$ is obtained to be compared to the
numerically measured value $0.63$.  The identity is also
supported by a constancy in the relative degree of roughening for
simulated fronts in stationary paper sheets
\cite{Lam}. 

\begin{figure}
   \psfull
   \centerline{\psfig{figure=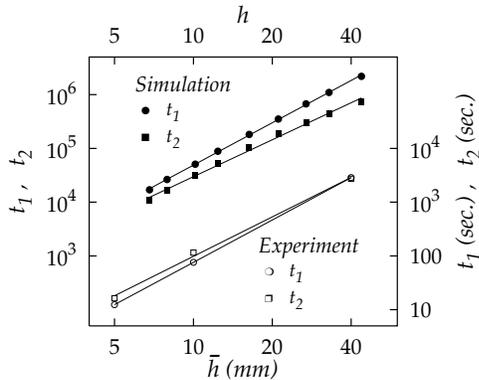,height=5truecm,clip=true}}
   \caption{\label{cgopt}\narrowtext
     Log-log plot of characteristic time $t_1$ and $t_2$ respectively for
flow and roughening against interface height $\bar{h}$ from experiment
(open symbols, right and bottom axes) and simulation (solid symbols,
left and top axes).
   }
\end{figure}

The unique time scale assumption and hence the identity are the main
objections by Dube et al. \cite{comment}.  We now give further
evidences by directly extracting relevant time constants from
experimental and numerical data in Refs. \cite{Horvath,Lam}. The mean
flow is characterized by $t_1 = \bar{h}/ v$. For roughening, we
observe that $C(t)$ can be nicely fitted by $\sqrt{2} w [ 1 +
(t/t_2)^{-a\beta}]^{-1/a}$ at $a=3.5$ and the relaxation time $t_2$
hence follows. Results are plotted in Fig. 1.  The unique time scale
assumption means that $t_1$ and $t_2$ are different representations of
a unique time scale for the front propagation dynamics. It requires
$t_1/t_2$ to be a constant of order, but $not$ necessarily exactly
unity.  In contrast, if flow and roughening were distinct dynamics
\cite{Dube}, $t_1$ and $t_2$ could easily differ by a factor of, say,
10 or 100.  Using the experimental results, $t_1/t_2 \approx 0.82$ on
average for various values of $\bar{h}$. It is close to unity and is a
strong experimental support of our view. A further numerical support
is that $t_1/t_2 \approx 2.0$ roughly from our simulation.  The very
similar magnitudes of $t_1$ and $t_2$ $both$ in experiment and
simulation independently, in addition to their approximate
proportionality, should not be causally dismissed as accidental.
Instead, its implication of a unique time scale can be a helpful hint
for a complete theory of the experimental observations.

The discrepancy between the ratios 0.82 and 2.0 deserves some caution,
though it can be due to finite size effects, which affects prefactors
more severely than exponents.  In Fig. 1, variations in the slopes of
the fitted lines for either the experimental or numerical values
reflect slight deviations from $t_1 \propto t_2$. These are
alternative representations of deviations of the estimated exponents
from our identity discussed above. They have been attributed to finite
size effects which are particularly significant in the simulations
\cite{Lam}.

C.H.L. is supported by HK RGC Grant 5191/99P.  V.K.H. is supported by
Hungarian Science Foundation grant OTKA-F17310, M-27950 and NATO grant
DGE-9804461.



\end{multicols}
\end{document}